\documentclass[aps,prl,superscriptaddress,showpacs,byrevtex,twocolumn]{revtex4-1}

\usepackage{color}

\usepackage{graphicx}
\usepackage{amsmath}
\usepackage{amsfonts}
\usepackage{amssymb}
\usepackage[latin1]{inputenc}
\begin{document}

\title{Observation of a molecule-metal interface charge transfer state by resonant photoelectron spectroscopy}
\date{\today}
\author{C. Sauer}
\author{M. Wie{\ss}ner}
\author{A. Sch\"oll}
\author{F. Reinert}
\affiliation{Universit\"at
W\"urzburg, Experimentelle Physik VII \& R\"ontgen Research Center
for Complex Material Systems RCCM, 97074 W\"urzburg, Germany}
\affiliation{Karlsruher Institut f\"ur Technologie KIT,
Gemeinschaftslabor f\"ur Nanoanalytik, 76021 Karlsruhe, Germany}


\begin{abstract}
We report the discovery of a novel charge transfer (CT) state at a molecule-metal interface by the application of resonant photoelectron spectroscopy (\mbox{ResPES}). This interface feature is neither present for molecular bulk samples nor for the clean substrate. Within a simplified two-step model this signal is assigned to a particular final state that is invisible in direct photoelectron spectroscopy but in \mbox{ResPES} revealed through relative resonant enhancement. A detailed analysis of the spectroscopic signature of the CT state shows characteristics of electronic interaction not found in other electron spectroscopic techniques. Our study demonstrates the sensitivity of \mbox{ResPES} to such interactions and constitutes a new way to investigate CT at molecule-metal interfaces.
\end{abstract}

\maketitle



One of the crucial questions for the performance of organic electronics is the charge transfer across the metal-organic interface. The technique of resonant photoelectron spectroscopy (\mbox{ResPES}) is able to investigate this issue within the core hole clock technique \cite{BruehwilerRevModPhys}. For small adsorbates like noble gases \cite{KarisPRL}, sulphur atoms \cite{FoehlischNature,FoehlischCPL} and small molecules \cite{FoehlischSurfSci} a quantitative extraction of charge transfer times has been successfully performed in the fs- and as-regime. Also for large $\pi$-conjugated molecules a rather large body of literature with a quantitative analysis of charge transfer times exist \cite{SchnadtNature,DeJongPRB,WeiApplPhysLett,CaoJourChemPhys}. However, the complicated electronic structure of these molecules and possible strong interactions with the substrate pose a tremendous challenge to the quantitative extraction of intensities from the \mbox{ResPES} data. Hence the published charge transfer times of similar systems differ substantially and the significance of many studies remains questionable. In order to permit quantitative investigations for these systems first the interaction at the interface and second its consequence for \mbox{ResPES} need to be understood in more detail. 

In \mbox{ResPES} certain signals in the \mbox{PES} spectrum can get enhanced due to additional autoionization (AI) channels which open up in the resonance case . For example the 6\,eV satellite in Ni metal \cite{HuefnerPhysLett} was assigned to a two hole final state due to the intensity enhancement and energy dispersion observed while tuning the photon energy ($h \nu$) over a resonance \cite{GuillotPRL,WeineltPRL}. Furthermore similar satellites for Cr and Fe metal could only be discovered due to resonant intensity enhancement \cite{HuefnerPRBCrFe}. In compounds $h \nu$ can be chosen to selectively excite in a resonance belonging to one of its constituents in order to enhance signals of this particular species \cite{HuefnerAdvPhys,DuoSurfSciRep,BernerPRL}. Moreover the surface sensitivity of \mbox{ResPES} allows to apply the concept of selective resonant enhancement to quasi two dimensional systems like surface alloys \cite{SchwabPRB}. However, which signal belonging to the selected species gets enhanced is a matter of localization of the resonantly excited electron. For the excitation into a delocalized d-band of a metal for example it is the incoherent Auger process that mainly gains in intensity and the enhanced signal disperses with a constant kinetic energy ($E_{K}$) \cite{GuillotPRL,WeineltPRL,HuefnerPRBCrFe,LopezZPhysB}. Exciting resonantly into a localized f-orbital on the other hand leads to a coherent and energy conserving process in which the enhanced signal stays at constant binding energy ($E_{B} = h \nu - E_{K}$) \cite{SchmidtMayPRB,MishraPRL,HuefnerPRBPrNd}. If a case of intermediate localization is realized both, the coherent and the incoherent signal will be visible. Directly on resonance the continuous \mbox{PES} channel and the discrete Auger channel interfere \cite{FanoPhysRev,DavisPRB} and cannot be distinguished but above the resonance the constant $E_{K}$ dispersion of the Auger signal makes this signal move away from the coherent signal at constant $E_{B}$. Consequently the coherent and the incoherent part can be separated for simple model systems \cite{KarisPRL}. For an adsorbate at a surface the latter can be interpreted as a charge transfer (CT) across the adsorbate substrate interface which in principle allows a determination of the CT time with the core hole clock technique \cite{BruehwilerRevModPhys}. So if the intensities of the signals corresponding to both channels can be extracted from the \mbox{ResPES} spectrum in a significant way a quantitative value of the CT time can be determined \cite{FoehlischNature,FoehlischCPL}.

\begin{figure*}
    \centering
        \includegraphics[width=0.75\textwidth]{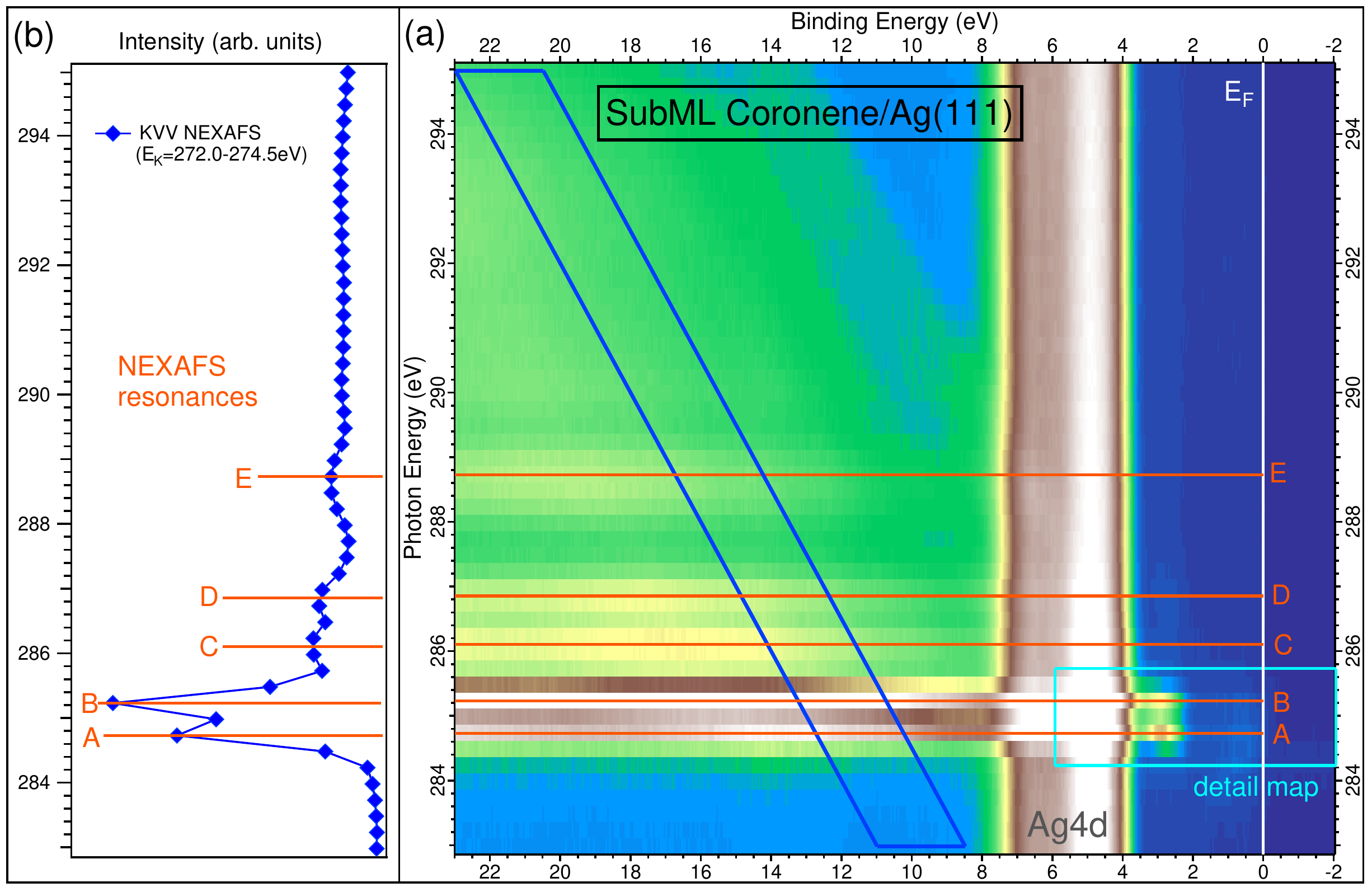}
    \caption{(color online). \mbox{ResPES} data of a SubML Coronene/Ag(111) recorded at the carbon K-edge.  \textbf{(a)} \mbox{PES} overview map with $h \nu$ increment of 250\,meV and $E_{B}$ increment of 50\,meV. The cyan box denotes the region measured for the \mbox{PES} detail map shown in Fig.\,\ref{fig:ResPESCoroDetailMap}(a). \textbf{(b)} \mbox{KVV NEXAFS} extracted from the blue box in the \mbox{PES} overview map (for details see text). The orange lines in (a) and (b) denote the \mbox{NEXAFS} resonances A - E.}
    \label{fig:ResPESCoroLargeMap}
\end{figure*}

In this work we apply \mbox{ResPES} to a model system of a large $\pi$-conjugated molecule adsorbed on a metal surface, namely a sub-monolayer (SubML) of Coronene adsorbed on Ag(111). First we present the observation of a CT state in a \mbox{PES} map recorded in the energetic region of the largest resonant enhancement of the highest occupied molecular orbital (\mbox{HOMO}). The CT state is found to originate from the metal-organic interface and the spectroscopic signature of this feature shows characteristics of strongly coupled molecule-metal interfaces. Hence \mbox{ResPES} reveals an electronic interaction between Coronene and Ag(111), a system that shows no evidence for strong coupling in other electron spectroscopic techniques. We then explain the appearance of the CT feature within a simplified two-step model that includes molecule-metal CT. Finally the absence of the CT state in direct \mbox{PES} is explained by relative resonant enhancement of this state with respect to the \mbox{HOMO}.

\begin{figure}
    \centering
        \includegraphics[width=0.38\textwidth]{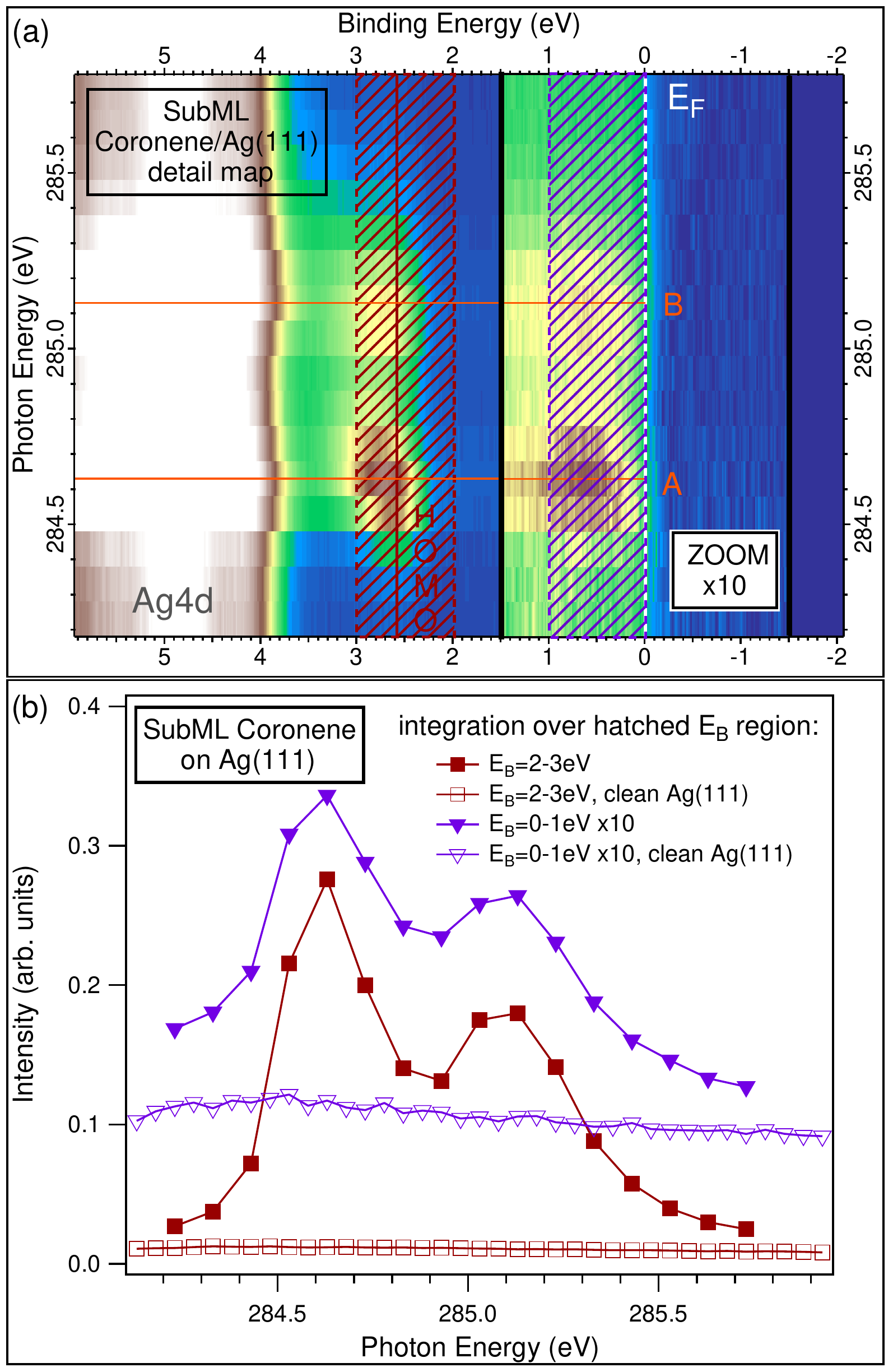}
    \caption{(color online). \textbf{(a)} \mbox{PES} detail map of a SubML Coronene/Ag(111) recorded at the carbon K-edge within the region denoted by the cyan box in Fig.\,\ref{fig:ResPESCoroLargeMap}(a) ($h \nu$ increment of 100\,meV, $E_{B}$ increment of 15\,meV). The $E_{B}$ region within the thick black lines ($E_{B} = -1.5$\,eV to $E_{B} = 1.5$\,eV) is multiplied by a factor of 10 with respect to the rest of the \mbox{PES} map. The orange horizontal lines denote the \mbox{NEXAFS} resonances A and B and the brown vertical line shows the $E_{B}$ of the \mbox{HOMO}. \textbf{(b)} Integrated intensities as a function of $h \nu$. Filled symbols stem form the hatched areas of the \mbox{PES} map displayed in panel (a), open symbols from a corresponding \mbox{PES} map recorded for clean Ag(111). The integrated intensities of $E_{B} = 0$\,eV$- 1$\,eV (triangles) are multiplied by a factor of 10 with respect to the integrated intensities of $E_{B} = 2$\,eV$- 3$\,eV (squares).}
    \label{fig:ResPESCoroDetailMap}
\end{figure}

Measurements were performed at BESSY II at the undulator beamline UE52-PGM ($E / \Delta E > 14000$ at 400\,eV photon energy, with cff=10 and 20\,$\mu$m exit slit \cite{RoccoJChemPhys}) in a UHV chamber with a pressure below $5*10^{-10}$\,mbar. All \mbox{PES} maps were recorded with p-polarized light and 70° angle of incidence with respect to the surface normal, a beamline exit slit of 40\,$\mu$m and a cff value of 10. Photoelectron intensities were detected with a Scienta R4000 electron analyzer with an energy resolution better than $\Delta E=35$\,meV for the \mbox{PES} detail map (Fig.\,\ref{fig:ResPESCoroDetailMap}(a)). $h \nu$ was calibrated with the Fermi edge ($E_{F}$) resulting in an accuracy better than 50\,meV (for $E_{B}$ and $h \nu$). \mbox{PES} intensities were normalized to the ring current and the beamline flux curve which was recorded separately by measuring the clean surface \cite{SchoelJourElSpec}. For the \mbox{PES} detail map (Fig.\,\ref{fig:ResPESCoroDetailMap}(a)) the 2nd order C1s signal was subtracted with a reference spectrum of the same sample previous to the normalization procedure. The Ag(111) substrate was cleaned by several sputter and annealing cycles and its cleanness was confirmed by \mbox{PES}. Coronene molecules were purified by sublimation and evaporated from a Knudsen cell at a pressure below $10^{-8}$\,mbar and at room temperature (RT). Film thickness was determined by core level intensities of the adsorbate and the substrate, using the effective electron attenuation lengths given in Ref.\,\cite{GraberSurfSci}.


Fig.\,\ref{fig:ResPESCoroLargeMap}(a) shows a \mbox{PES} overview map at the carbon K-edge of a SubML Coronene/Ag(111). The most prominent signals are the Ag4d bands in between a $E_{B}$ of 4\,eV and 8\,eV. Below the onset of the first absorption peak of Coronene at approximately 284\,eV these are the only significant contribution to the spectra. At higher $h \nu$ (but still below the direct photoionization into vacuum) the situation changes dramatically and signals originating from the Coronene get massively enhanced. At higher $E_{B}$ than the Ag4d bands both, low lying molecular orbitals (MO) \cite{WiessnerNJPhys} and \mbox{KVV Auger} peaks, are situated. The former can be seen as enhanced \mbox{PES} signals due to AI after photon absorption while the latter are a consequence of core hole decays. Integrating over a constant $E_{K}$ from 272.0\,eV to 274.5\,eV denoted by the blue box in Fig.\,\ref{fig:ResPESCoroLargeMap}(a) results in the spectrum displayed in Fig.\,\ref{fig:ResPESCoroLargeMap}(b). This spectrum is equal (within the chosen $h \nu$ increment) to the partial electron yield near edge X-ray absorption fine structure spectroscopy (\mbox{PEY NEXAFS}) spectrum but only consists of a certain $E_{K}$ window instead of all emitted electrons above a retarding voltage as for the \mbox{PEY NEXAFS}. Since its signal mainly originates from several \mbox{KVV Auger} decays we call it \mbox{KVV NEXAFS} (for further details see Ref.\,\cite{SauerResPESElVib}). In the $h \nu$ region of two most intense \mbox{NEXAFS} resonances A and B we further observe a substantial intensity enhancement of the \mbox{HOMO} which is located at slightly lower $E_{B}$ than the rising edge of the Ag4d bands. For closer inspection of these signals the area within the cyan box in Fig.\,\ref{fig:ResPESCoroLargeMap}(a) is recorded in a subsequent measurement with lower energy increments and higher resolution.

Fig.\,\ref{fig:ResPESCoroDetailMap}(a) displays this \mbox{PES} detail map. Here not only the intensity enhancement of the \mbox{HOMO} but also its line shape variation as a function of $h \nu$ becomes obvious. This effect is due to a difference in the vibrational progression of the \mbox{HOMO} which is a consequence of the particular vibronic excitations within the photon absorption \cite{SchoellPRL} and hence a function of $h \nu$ (for a detailed discussion see Ref.\,\cite{SauerResPESElVib}). The focus of this work lies on the additional feature centered at approximately 0.5\,eV $E_{B}$ below $E_{F}$ which becomes clearly visible after the intensity of the \mbox{PES} detail map is multiplied by a factor of 10 within the thick black lines. Integrating over the constant $E_{B}$ windows marked by the hatched areas in Fig.\,\ref{fig:ResPESCoroDetailMap}(a) results in the filled symbols displayed in Fig.\,\ref{fig:ResPESCoroDetailMap}(b). Here it is revealed that the \mbox{HOMO} and the low $E_{B}$ feature exhibit a very similar intensity variation as a function of $h \nu$ and that the relative intensity of the low $E_{B}$ feature with respect to the \mbox{HOMO} intensity is found to be approximately 5\%. An integration over the same $E_{B}$ region of a corresponding \mbox{PES} detail map of the clean Ag(111) substrate (open symbols in Fig.\,\ref{fig:ResPESCoroDetailMap}(b)) shows no dependency on $h \nu$ even close to the magnitude observed for the SubML Coronene/Ag(111) sample. Furthermore the low $E_{B}$ feature is not observed in the corresponding \mbox{PES} data of a Coronene multilayer film \cite{SauerResPESElVib}. This leads to the conclusion that the feature originates from the metal-organic interface since neither the clean Ag(111) nor the pure Coronene film (without interface contribution) exhibits such a signal. Hence an interaction of the Ag(111) substrate with the Coronene adsorbate film must be present.

\begin{figure}
    \centering
        \includegraphics[width=0.38\textwidth]{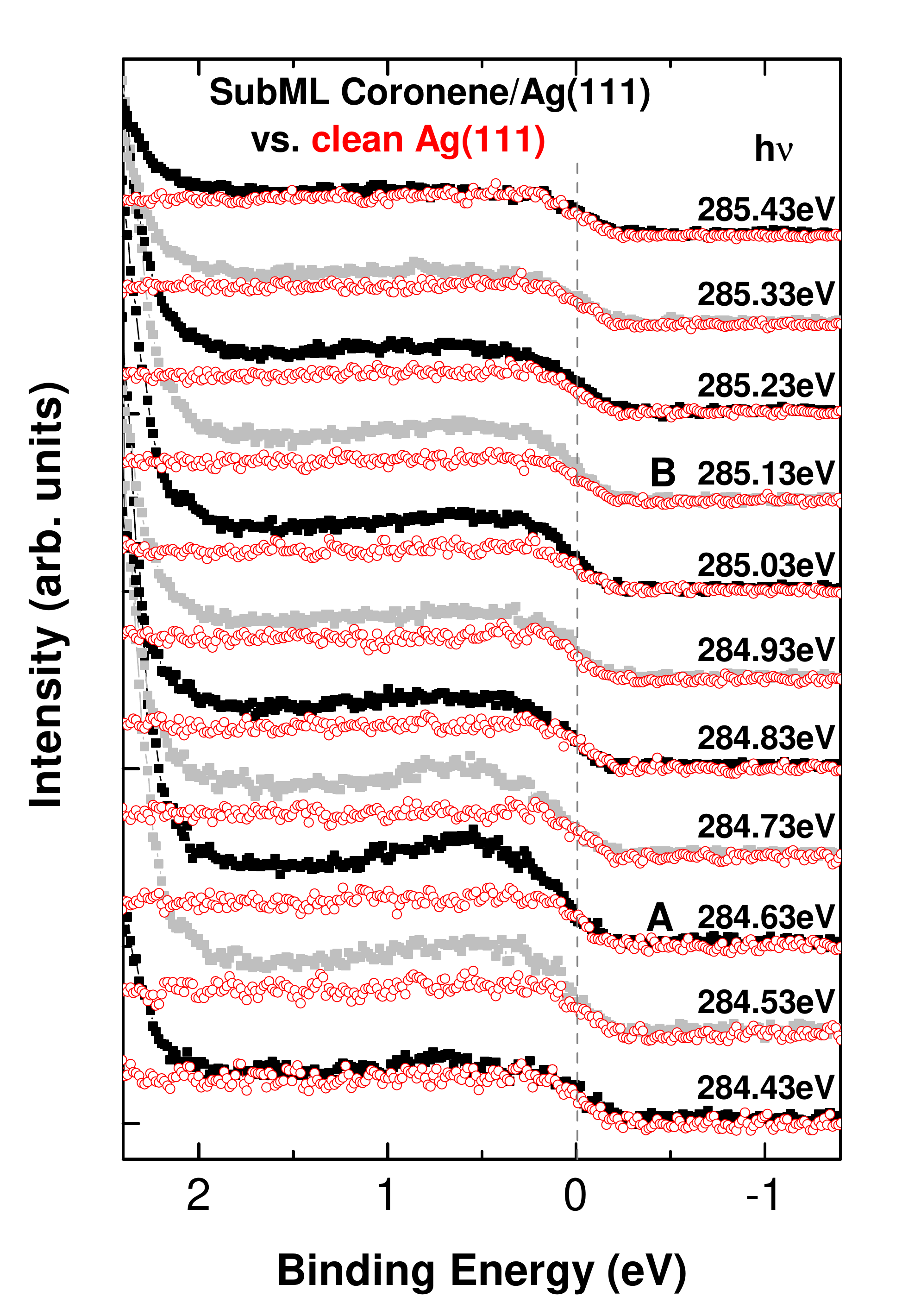}
    \caption{(color online). EDCs from the \mbox{PES} detail map of a SubML Coronene/Ag(111) displayed in Fig.\,\ref{fig:ResPESCoroDetailMap}(a) (black/grey) in comparison to corresponding EDCs from a \mbox{PES} detail map of clean Ag(111) (red). All EDCs from the \mbox{PES} detail map of the clean Ag(111) were multiplied by the same factor in order to match the Fermi edges of the uppermost EDCs ($h \nu = 285.43eV$).}
    \label{fig:ResPESCoroWaterfall}
\end{figure}

This conclusion is corroborated by the broad line shape of the low $E_{B}$ feature which is presented in Fig.\,\ref{fig:ResPESCoroWaterfall}. Here energy distribution curves (EDCs) from the \mbox{PES} detail map (Fig.\,\ref{fig:ResPESCoroDetailMap}(a)) are compared to EDCs from the corresponding \mbox{PES} detail map of clean Ag(111) in the relevant $E_{B}$ region. The broad and smeared out resonantly enhanced intensity of the SubML Coronene/Ag(111) film is similar to the signals observed for strongly coupled molecule-metal interfaces \cite{BendounanSurfSci} which are characterized by an occupied lowest unoccupied molecular orbital (\mbox{LUMO}) \cite{ZouSurfSci}. Interestingly, in direct \mbox{PES} (with $h \nu$ below the resonance) a possible \mbox{LUMO} signal is found to be below the detection limit of approximately 1\% with respect to the \mbox{HOMO} intensity at any point in the probed $k$-space region \cite{WiessnerNJPhys}. Moreover, the generally observed strong variation of \mbox{PES} core level and \mbox{NEXAFS} spectra of the strongly coupled molecule-metal systems with respect to the multilayer spectra cannot be found for Coronene. The only hint for an electronic interaction is a broadening of the single vibronic components of the \mbox{HOMO} in the high resolution \mbox{PES} spectrum of 1\,ML Coronene/Ag(111) with respect to 1\,ML Coronene/Au(111) \cite{WiessnerDiss}. Thus the finding of electronic interaction at the here investigated molecule-metal interface demonstrates the sensitivity of \mbox{ResPES} to such interactions.

An explanation for the observed low $E_{B}$ feature can be given on the basis of a simplified two-step picture (step (a) and (b)) including CT from the metal to the molecule. The \mbox{NEXAFS} resonances A and B can be assigned to an excitation into the \mbox{LUMO} \cite{OjiJChemPhys} thus this constitutes step (a) of the mechanism. Step (b) is the subsequent AI process. The channel leading to the enhanced \mbox{HOMO} signal (denoted as channel (I)) can be written in the following way:
\begin{align*}
	\text{(Ia)} \;\; & \left|C^2 H^2 L^0 M^{n}\right\rangle + h \nu \xrightarrow[no\:CT]{NEXAFS} \left|C^1 H^2 L^1 M^{n}\right\rangle \\
	\text{(Ib)} \;\; & \left|C^1 H^2 L^1 M^{n}\right\rangle \xrightarrow[]{AI} \left|C^2 H^1 L^0 M^{n}\right\rangle + e^- \\
\end{align*}
Hereby $C$ stands for the C1s core level, $H$ for the \mbox{HOMO}, $L$ for the \mbox{LUMO}, $M$ for the metal and $e^-$ is the emitted electron. All other levels of the system are omitted. The used nomenclature for the wave function of the molecule in contact to a metal is based on the cluster model description of transition metal oxides \cite{SawatzkyPRL,ZaanenPRL,FujimoriPRBR,FujimoriPRB,KotaniProgTheoPhys,ParlebasJourElSpec}. The superscript denotes the occupation of the particular level. In the same way we can write down the mechanism including CT: 
\begin{align*}
	\text{(IIa)} \;\; & \left|C^2 H^2 L^0 M^{n}\right\rangle + h \nu \xrightarrow[CT]{NEXAFS} \left|C^1 H^2 L^2 M^{n-1}\right\rangle \\
	\text{(IIb1)} \;\; & \left|C^1 H^2 L^2 M^{n-1}\right\rangle \xrightarrow[]{AI} \left|C^2 H^1 L^1 M^{n-1}\right\rangle + e^- \\ 
	\text{(IIb2)} \;\; & \left|C^1 H^2 L^2 M^{n-1}\right\rangle \xrightarrow[]{AI} \left|C^2 H^2 L^0 M^{n-1}\right\rangle + e^- \\
\end{align*}
This channel (denoted as channel II) can lead to two different final states ((1) and (2)) through AI which both could lead to the low $E_{B}$ feature in \mbox{ResPES}. Due to the involvement of interface CT in the mechanism leading to this state we will call it CT state. The fact that this CT state is not observed in direct \mbox{PES} requires a relative enhancement of this state with respect to the \mbox{HOMO} in \mbox{PES} at the K-edge. In other words the probability for the responsible step must be larger than the one for step (Ib). For the steps (IIb1) and (Ib) equal states, namely $C$, $H$ and $L$, are involved in the AI process so a different probability is highly unlikely. In contrast, step (IIb2) involves $C$ and the state $L$ twice. It is reasonable to expect the overlap matrix element of $L$ with itself to be larger than the ones of (Ib) and (IIb1) in which an overlap of $L$ and $H$ is included. Thus we assign the final state $\left|C^2 H^2 L^0 M^{n-1}\right\rangle$ to the observed CT state in \mbox{ResPES}. In principle this final state could also appear in a direct \mbox{PES} measurement as long as the coupling between the substrate states and the MO is nonvanishing. The probability for this state to be produced is apparently unlikely and hence it can only be detected through the resonant enhancement due to the AI channel in \mbox{ResPES}.

An equivalent and quantitative description of the process responsible for the valence satellite can be conducted on the basis of the Cluster Model along the lines of Ref.\,\cite{FujimoriPRB}. Here the involved states are represented as a quantummechanical superposition of a state with CT and another one without CT. A consistent description of the SubML Coronene/Ag(111) data with physically reasonable Cluster Model parameters requires a weighting factor of only $\frac{\text{(IIb2)}}{\text{(Ib)}} = 6$ in order to reproduce the observed relative resonant enhancement from \mbox{PES} to \mbox{ResPES}. This ratio of two overlap matrix elements containing different MO seems to be of reasonable magnitude. Details will be given in a future publication.


In conclusion the finding of a CT state in \mbox{ResPES} applied to a molecule-metal interface provides evidence for interfacial CT and electronic interaction not deduced by results from other electron spectroscopic techniques. Through the description of the responsible mechanism within a simple two-step model we were able to assign a particular final state to this CT state. The found process requires the involvement of metal states and CT into the molecule. In addition to manifesting the power of \mbox{ResPES} in revealing electronic interaction at interfaces, our experiments demonstrate a novel route to identify and characterize charge transfer processes at molecule-metal interfaces.

\acknowledgments{We thank the BESSY staff for support during beamtimes. Further we would like to thank M. Mulazzi, H. Schwab and F. Meyer for stimulating discussions. This work was supported by BESSY, by the Bundesministerium f\"ur Bildung und Forschung BMBF (grant no. 05K10WW2 and 03SF0356B) and by the Deutsche Forschungsgemeinschaft DFG (GRK 1221 and RE1469/9-1).}

\end{document}